\documentclass{INTERSPEECH2023}

\interspeechcameraready

\usepackage{hyperref}
\usepackage{siunitx}
\usepackage{mathtools}
\usepackage{comment}
\usepackage{enumitem}
\usepackage{threeparttable}
\usepackage{multirow}
\usepackage{subfig}

\usepackage{xspace}
\makeatletter
\DeclareRobustCommand\onedot{\futurelet\@let@token\@onedot}
\def\@onedot{\ifx\@let@token.\else.\null\fi\xspace}

\def\ie{\emph{i.e}\onedot}

\makeatother

\def\equationautorefname~#1\null{(#1\null)}

\newcommand{\vect}[1]{{\mbox{\boldmath $#1$}}}

\def\appendixautorefname~#1\null{~#1 \null}

\title{Spoofing Attacker Also Benefits from Self-Supervised Pretrained Model}
\name{
    Aoi Ito$^1$\sthanks{\enspace The authors equally contributed to this work. This work has been done during Aoi Ito's internship at Hitachi, Ltd.},
    Shota Horiguchi$^{2*}$
}

\address{
  $^1$ Hosei University,
  $^2$ Hitachi, Ltd.
}
\email{aoi.ito.8q@stu.hosei.ac.jp, shota.horiguchi.wk@hitachi.com}

\begin{document}
\abovedisplayskip=4pt
\belowdisplayskip=4pt
\setlength\textfloatsep{11pt}
\setlength\abovecaptionskip{5pt}
\setlength\dbltextfloatsep{9pt}

\maketitle
\begin{abstract}
Large-scale pretrained models using self-supervised learning have reportedly improved the performance of speech anti-spoofing.
However, the attacker side may also make use of such models.
Also, since it is very expensive to train such models from scratch, pretrained models on the Internet are often used, but the attacker and defender may possibly use the same pretrained model.
This paper investigates whether the improvement in anti-spoofing with pretrained models holds under the condition that the models are available to attackers.
As the attacker, we train a model that enhances spoofed utterances so that the speaker embedding extractor based on the pretrained models cannot distinguish between bona fide and spoofed utterances.
Experimental results show that the gains the anti-spoofing models obtained by using the pretrained models almost disappear if the attacker also makes use of the pretrained models.

\end{abstract}

\section{Introduction}
Automatic speaker verification (ASV) is becoming a possible choice for secure authentication with the recent progress in its performance.
However, ASV systems are exposed to the menaces of malicious attacks using speech synthesis, voice conversion, and replaying of recorded voice.
To protect systems from such attacks, anti-spoofing methods are widely studied along with the growth of the community led by the ASVspoof challenges \cite{wu2017asvspoof}.
Although the boundaries are vague due to end-to-end modeling \cite{tak2021end}, anti-spoofing models generally consist of a feature extraction part and a classification part.
As the input feature, hand-crafted features such as linear frequency cepstral coefficients \cite{alegre2013one} and constant Q cepstral coefficients \cite{todisco2017constant} or features from neural networks such as SENet \cite{lai2019assert}, DenseNet \cite{zhang2020improving}, and RawNet2 \cite{tak2021end} are used in the literature.

The quality of features extracted from audio using neural networks is rapidly advancing with the self-supervised learning (SSL) paradigm; a lot of models have been proposed in the last few years such as wav2vec 2.0 \cite{baevski2020wav2vec}, HuBERT \cite{hsu2021hubert}, and WavLM \cite{chen2022wavlm}.
They have shown greatly improved performance on various speech-related tasks such as speech recognition \cite{baevski2020effectiveness}, speech enhancement \cite{huang2022investigating}, speaker identification \cite{fan2021exploring}, and emotion recognition \cite{pepino2021emotion}.
Likewise, it has been reported that ASV anti-spoofing can also benefit from SSL models \cite{xie2021siamese,wang2022investigating,tak2022automatic}.

Although SSL models are powerful, training them from scratch consumes a lot of computational resources.
For example, wav2vec 2.0, HuBERT, and WavLM have reportedly been trained using 64, 32, and 16 NVIDIA V100 GPUs, respectively, even for the smallest \textsc{Base} model of each.
These computing environments are not necessarily on a scale that is readily available to everyone.
Therefore, it is a common practice to finetune publicly available pretrained models on the Internet when using them for one's own applications.

\begin{figure}[t]
    \includegraphics[width=\linewidth]{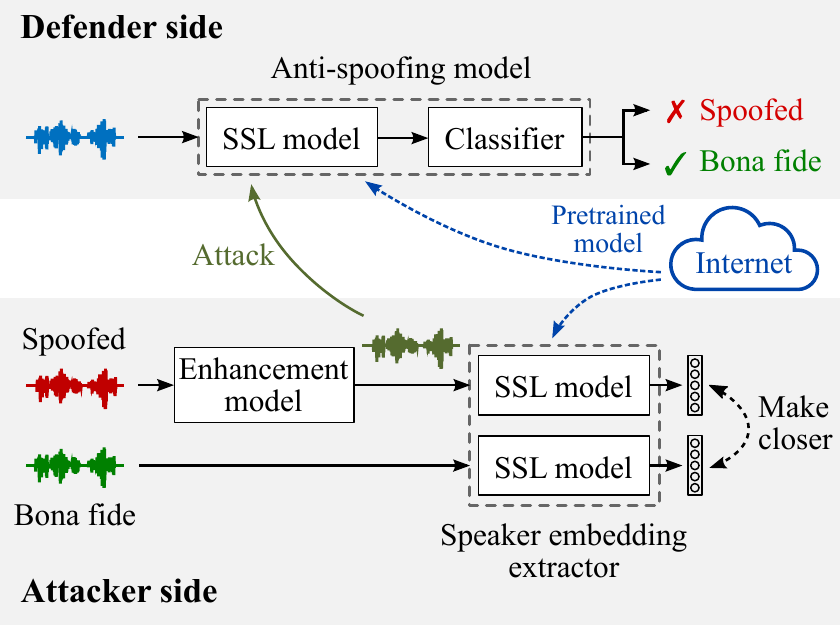}
    \caption{Overview of evaluation framework}
    \label{fig:overview}
\end{figure}

Here, although the accuracy of ASV anti-spoofing has indeed been reportedly improved by SSL models, it is easy to imagine that the attacker side can also take advantage of the power of these models.
There is no previous investigation of the extent to which anti-spoofing loses the gains obtained by SSL models when attackers also use them.
Also, if an ASV anti-spoofing system is developed using a publicly available pretrained model, an attacker can also access the same model.
The security risks from the same pretrained models being used in attacks against anti-spoofing systems have also not been explored before.

This paper aims to provide answers to two questions: i) if an attacker leverages SSL models, will they maintain the anti-spoofing performance reportedly improved by them, and ii) is there any disadvantage to anti-spoofing if the attacker uses the same pretrained model on the Internet as the defender even when it is finetuned?
For this purpose, we propose a method for training the attacker, as illustrated in \autoref{fig:overview}.
In the method, a speaker embedding extractor based on SSL models is first trained.
Then, an enhancement model that improves the deception ability of spoofed utterances is trained to make the speaker embeddings extracted from bona fide and \textit{enhanced} spoofed utterances closer.
The proposed method is evaluated using various combinations of attacker-defender pairs on three ASVspoof challenge datasets.

\section{Related work}
If the attacker has prior knowledge of an anti-spoofing model, adversarial attacks can be attempted: white-box attacks when the parameters of the model are known \cite{szegedy2013intriguing} and black-box attacks when only input-output pairs of the model are known \cite{papernot2017practical} (please refer to \cite{das2020attacker} for a literature review).
The performance of ASV with SSL models was also degraded by adversarial attacks \cite{wu2022characterizing}, but the models were not finetuned.
When a model is fine-tuned, it is not possible to directly calculate such adversarial attacks.
One possible attack in such a case is when a pretrained model has a backdoor in it due to weight poisoning \cite{kurita2020weight}, but this is not considered in this paper because we assume the commonly used SSL model.
Another possibility is to utilize the transferability of adversarial attacks; successful attacks on a certain model are also likely to fool other models \cite{zhang2020black}.
While attacks by adversarial samples are an important issue for any machine learning model, the scope of this paper is to verify whether the attacker can also benefit from the power of SSL models, so generating such samples artificially is out of scope for this paper.

The paradigm used in this paper is highly related to verification-to-synthesis (V2S) \cite{nakamurav2s}, in which a voice conversion model is trained to fool a pretrained speaker classification model.
There are three differences from V2S: i) we enhance utterances that are already made to spoof target speakers, ii) we do not apply any regularization to enhanced utterances to preserve their phonetic properties, and iii) we directly compare a pair of speaker embeddings instead of the output from the classifier and the desired speaker label, which makes the method applicable to any speaker.

\section{Method}
We assume the situation depicted in \autoref{fig:overview}.
The defender side builds an anti-spoofing model by using an SSL model on the Internet as a frontend.
The anti-spoofing model performs a two-way classification that distinguishes whether an input is a bona fide or spoofed utterance.
The attacker side aims to obtain a spoofing model that transforms an already spoofed piece of audio by, for example, voice conversion or text-to-speech synthesis, to enhance its spoofing ability.
With a speaker embedding extractor based on the SSL model, the spoofing model is trained so that the embedding extracted from a spoofed input is indistinguishable from that from a bona fide input.
We will describe the implementation of each side in detail in the following subsections.

\subsection{Defender side}\label{sec:defender}
For the anti-spoofing model, we used a simplified version of the RawNet2-based architecture \cite{tak2022automatic}, in which the SincNet frontend \cite{ravanelli2018speaker} was replaced by an SSL model.
\autoref{tbl:anti_spoofing_model} shows the detailed configuration of the model.
Given 64,600 samples of audio ($\sim$4 seconds), the SSL frontend first extracts 201-length 768-dimensional speech representations.
The following dimensionality reduction layer and six-stacked residual blocks further convert the representations to perform two-way classification, \ie, spoofed vs. bona fide, in the last layer.
The model was trained using the standard cross entropy loss.
Note that the parameters of the SSL frontend were initialized with those of the pretrained model and jointly optimized with the backend in an end-to-end manner.

\begin{table}[t]
    \centering
    \caption{Architecture of anti-spoofing model. Input to model is 64600-length waveform. Output shape corresponds to number of channels, frames, and frequency bins, respectively. BN and SeLU refer to batch normalization and scaled exponential linear unit, respectively.}
    \label{tbl:anti_spoofing_model}
    \resizebox{\linewidth}{!}{%
    \begin{tabular}{@{}lcl@{}}
    \toprule
    Layer& Output shape&Configuration\\\midrule\midrule
    SSL frontend&(1, 201, 768)&wav2vec 2.0/HuBERT/WavLM\\\midrule
    \multirow{3}[3]{*}{\shortstack[l]{Dimensionality\\reduction}}&(1, 201, 128)&128-dim fully connected\\\cmidrule(l){2-3}
    &(1, 67, 42)&$3\times 3$ max pooling\\\cmidrule(l){2-3}
    &(1, 67, 42)&BN \& SeLU\\\midrule
    Residual block&(32, 67, 42)&$\begin{bmatrix}\text{3x3 conv, 32-ch}\\\text{BN \& SeLU}\\\text{3x3 conv, 32-ch}\end{bmatrix}\times 2$\\\midrule
    Residual block&(64, 67, 42)&$\begin{bmatrix}\text{3x3 conv, 64-ch}\\\text{BN \& SeLU}\\\text{3x3 conv, 64-ch}\end{bmatrix}\times 4$\\\midrule
    \multirow{2}[2]{*}{Classification}&64&$67\times 42$ global average pooling\\\cmidrule(l){2-3}
    &2&2-dim fully connected\\
    \bottomrule
    \end{tabular}%
    }
\end{table}

\subsection{Attacker side}
The attacker side attempts to transform spoofed recordings into ones that the SSL model cannot distinguish from bona fide recordings.
The training of the spoofing model is two-staged.
First, a speaker embedding extractor is constructed on the basis of the model pretrained using SSL, and then an enhancement model to improve the input's spoofing ability is trained using the extractor.
If the enhancement model can be trained to be able to fool the extractor, then an anti-spoofing model based on the same SSL model could be fooled as well.

The speaker embedding extractor $f_\text{embed}$ was trained to classify utterances on the basis of their speaker IDs.
Given an input utterance, the SSL frontend first computes frame-level embeddings, which are then aggregated by average pooling along the time axis to obtain an utterance-level embedding.
Following the conventional study \cite{vaessen2022finetuning}, the entire network was optimized using additive angular margin (AAM) softmax loss \cite{deng2022arcface}.

With the well-trained speaker embedding extractor, the enhancement model $f_\text{enh}$ was trained to convert spoofed utterances to ones whose speaker embeddings are not distinguishable from those extracted from bona fide utterances.
During training, a pair of bona fide and spoofed utterances is used to train the model.
A spoofed utterance $\vect{x}_\text{spoof}$ is first fed to the spoofing model to convert it to an enhanced one:
\begin{equation}
    \vect{x}_\text{enh}=f_\text{enh}\left(\vect{x}_\text{spoof}\right).\label{eq:enhance}
\end{equation}
For the enhancement model ${f_\text{enh}}$, we used Conv-TasNet \cite{luo2019conv} to convert the input audio in the time domain. 
Then, speaker embeddings are extracted from each bona fide and \textit{enhanced} spoofed utterances.
The network is optimized to minimize the angle between those embeddings by using the following loss:
\begin{equation}
    \mathcal{L}=1-\cos\left(f_\text{embed}\left(\vect{x}_\text{enh}\right),f_\text{embed}\left(\vect{x}_\text{bonafide}\right)\right),\label{eq:loss}
\end{equation}
where $\vect{x}_\text{bonafide}$ is a bona fide utterance of the speakers who the spoofed utterance $\vect{x}_\text{spoof}$ is pretending to be and $\cos\left(\cdot,\cdot\right)$ denotes the cosine similarity between two arguments.
Note that the parameters of the speaker embedding extractor were frozen during the training of the enhancement model; otherwise, it will fall into a trivial solution that, for example, always outputs the same embedding regardless of the input.

During evaluations, \textit{enhanced} spoofed utterances obtained using \autoref{eq:enhance} are fed to the anti-spoofing models described in \autoref{sec:defender} instead of the original spoofed utterances.

\section{Experimental settings}
\subsection{Dataset}
We consider two scenarios in this paper.
The first scenario is that an attacker and a defender use a different dataset for the training of each side's model.
To meet this purpose, each of the training and development sets of the ASVspoof 2019 logical access (LA) database \cite{todisco2019asvspoof,wang2020asvspoof} was divided into two portions to train spoofing models and anti-spoofing models, respectively.
For the bona fide utterances, we assigned half of each speaker's utterances to the attacker and the other half to the defender.
For the spoofed utterances, assuming that an attacker does not have prior knowledge of the specific method that a defender is taking into consideration, we divided them based on their systems: A01, A03, A05 for the attacker and A02, A04, A06 for the defender\footnote{This split was to avoid giving unfair advantages to the attacker since the spoofing systems used for A04 and A06 were also used in the test set.}.
The second scenario is that an attacker has access to some of the defender's data, e.g., the defender makes use of publicly available datasets.
In this scenario, the defender uses the whole ASVspoof 2019 LA database, while the attacker uses the same portion as the first scenario.

For the evaluation, we used the test set of the ASVspoof 2019 LA database as a clean dataset, in which spoofing attacks are based on speech synthesis and voice conversion.
We also used the ASVspoof 2021 LA and deepfake (DF) databases \cite{yamagishi2021asvspoof} for more noisy and realistic trials.
The 2021 LA database is based on the same attack algorithms as the 2019 LA database, but the effects of encoding and transmission over the telephone are also taken into account. The 2021 DF database focuses on the distortion caused by compressing and restoring audio through various codecs.
Although there is some discussion about the appropriateness of the original ASVspoof datasets \cite{muller2021speech}, we used them as they are.
The models' performance was evaluated using an equal error rate (EER).

\subsection{Model configuration}
As the SSL models, we used wav2vec 2.0 \textsc{Base} \cite{baevski2020wav2vec}, HuBERT \textsc{Base} \cite{hsu2021hubert}, WavLM \textsc{Base}, and WavLM \textsc{Base+} \cite{chen2022wavlm} models.
For simplicity, the word ``\textsc{Base}'' will be omitted hereafter.
The \texttt{TorchAudio} \cite{yang2022torchaudio} implementations of wav2vec 2.0 and HuBERT and the official implementation of WavLMs\footnote{\url{https://github.com/microsoft/unilm/tree/master/wavlm}} were used in our experiments.
Each model has approximately 95 million parameters.
The wav2vec2.0, HuBERT, and WavLM were trained with the concatenation of \textit{train-clean-100}, \textit{train-clean-360}, and \textit{train-other-500} from the LibriSpeech dataset \cite{panayotov2015librispeech}, and WavLM+ was trained with the Libri-Light \cite{kahn2020libri}, GigaSpeech \cite{chen2021gigaspeech}, and VoxPopuli \cite{wang2021voxpopuli} datasets.

The anti-spoofing models were trained on the four types of SSL models above.
They were trained using the Adam optimizer \cite{kingma2015adam} with a fixed learning rate of $1\times10^{-6}$ for at most 100 epochs.
The batch size was set to 32.
As the baseline without an SSL frontend, we also used the official RawNet2 recipe from the ASVspoof 2021 baseline\footnote{\url{https://github.com/asvspoof-challenge/2021}}.
No data augmentation techniques were applied during training.
The inputs to each model were aligned to 64,600 samples by cropping and/or repeating the original utterances.

For speaker embedding extractors, we simply finetuned the wav2vec 2.0 and HuBERT models using VoxCeleb2 \cite{nagrani2020voxceleb} and evaluated them using VoxCeleb1.
No extra embedding layer as the backend was introduced; thus, the dimensionality of the speaker embedding was 768.
Each model was optimized to minimize the AAM softmax loss with a margin of 0.3 and a scale of 15 using the Adam optimizer for 6 epochs ($\sim$100k iterations).
The learning rate was linearly increased from zero to $1\times 10^{-5}$ for the first \SI{10}{\percent} of iterations, kept unchanged for the next \SI{40}{\percent} of iterations, and then linearly decayed to zero for the final \SI{50}{\percent} of iterations.
Here also, we did not apply any data augmentation techniques during training.

The spoofing models based on Conv-TasNet were trained using each speaker embedding extractor as a backend.
We used a Conv-TasNet model that consists of three repetitions of eight-stacked convolutional blocks with different dilations, which was the best configuration in the original paper \cite{luo2019conv}.
The training was conducted for 300 epochs using the Adam optimizer with a fixed learning rate of $1\times10^{-5}$ and a batch size of 8 without data augmentation.

\section{Results}
\subsection{Preliminary results on attacker side}
Before discussing the main results, we report the performance of the model trained for the attacker side.
\autoref{tbl:speaker_identification} shows the EERs of the speaker embedding extractors on the VoxCeleb1 dataset.
We use three cleaned splits of the dataset: the original test set (VoxCeleb1 in \autoref{tbl:speaker_identification}), the extended test set (VoxCeleb1-E), and the hard test set (VoxCeleb-H).
Although the quality of the speaker embeddings is not the main focus of this paper, we note that these values are comparable, though not the best, to previously reported values \cite{vaessen2022finetuning}.

\begin{table}
    \centering
    \caption{Speaker verification results on VoxCeleb1 dataset}
    \label{tbl:speaker_identification}
    \scalebox{0.96}{%
    \begin{tabular}{@{}lccc@{}}
        \toprule
        &\multicolumn{3}{c}{EER (\%)}\\\cmidrule(l){2-4}
        SSL model&VoxCeleb1&VoxCeleb1-E&VoxCeleb1-H\\\midrule
        wav2vec 2.0&2.49&2.91&6.05\\
        HuBERT& 2.85&3.15&6.49\\
        \bottomrule
    \end{tabular}%
    }
\end{table}

\subsection{Main results}

\begin{table*}[t]
    \centering
    \caption{EERs (\%) when anti-spoofing models were trained using whole ASVspoof 2019 LA training set.}
    \label{tbl:main_results_all}
    \setlength{\tabcolsep}{3pt}
    \subfloat[ASVspoof 2019 LA test set\label{tbl:main_results_all_2019LA}]{%
    \resizebox{0.32\linewidth}{!}{%
    \begin{tabular}{@{}lccc@{}}
        \toprule
        &\multicolumn{3}{c}{Spoofing enhancement}\\\cmidrule(l){2-4}
        Anti-spoofing&None&wav2vec 2.0&HuBERT\\\midrule
        RawNet2&17.49&60.60&57.03\\
        wav2vec 2.0&0.81&0.60&0.71\\
        HuBERT&1.62&2.86&2.83\\
        WavLM&1.03&2.68&2.53\\
        WavLM+&0.44&0.24&0.23\\
        \bottomrule
    \end{tabular}%
    }}\hfill
    \subfloat[ASVspoof 2021 LA test set\label{tbl:main_results_all_2021LA}]{%
    \resizebox{0.32\linewidth}{!}{%
    \begin{tabular}{@{}lccc@{}}
        \toprule
        &\multicolumn{3}{c}{Spoofing enhancement}\\\cmidrule(l){2-4}
        Anti-spoofing&None&wav2vec 2.0&HuBERT\\\midrule
        RawNet2&18.06&67.68&63.70\\
        wav2vec 2.0&7.20&16.57&16.47\\
        HuBERT&4.89&25.94&23.95\\
        WavLM&7.99&33.95&32.59\\
        WavLM+&7.55&26.94&25.66\\
        \bottomrule
    \end{tabular}%
    }}\hfill
    \subfloat[ASVspoof 2021 DF test set\label{tbl:main_results_all_2021DF}]{%
    \resizebox{0.32\linewidth}{!}{%
    \begin{tabular}{@{}lccc@{}}
        \toprule
        &\multicolumn{3}{c}{Spoofing enhancement}\\\cmidrule(l){2-4}
        Anti-spoofing&None&wav2vec 2.0&HuBERT\\\midrule
        RawNet2&24.01&68.46&65.77\\
        wav2vec 2.0&10.31&25.49&26.36\\
        HuBERT&18.93&46.22&45.49\\
        WavLM&16.14&46.71&45.97\\
        WavLM+&11.08&33.63&32.24\\
        \bottomrule
    \end{tabular}%
    }}
\end{table*}

\begin{table*}[t]
    \centering
    \caption{EERs (\%) when anti-spoofing models were trained using portion of ASVspoof 2019 LA training set.}
    \label{tbl:main_results_part}
    \setlength{\tabcolsep}{3pt}
    \subfloat[ASVspoof 2019 LA test set\label{tbl:main_results_part_2019LA}]{%
    \resizebox{0.32\linewidth}{!}{%
    \begin{tabular}{@{}lccc@{}}
        \toprule
        &\multicolumn{3}{c}{Spoofing enhancement}\\\cmidrule(l){2-4}
        Anti-spoofing&None&wav2vec 2.0&HuBERT\\\midrule
        RawNet2&10.93&24.28&24.62\\
        wav2vec 2.0&0.82&0.60&0.67\\
        HuBERT&1.70&2.12&2.09\\
        WavLM&0.94&3.05&2.95\\
        WavLM+&0.73&0.30&0.28\\
        \bottomrule
    \end{tabular}%
    }}\hfill
    \subfloat[ASVspoof 2021 LA test set\label{tbl:main_results_part_2021LA}]{%
    \resizebox{0.32\linewidth}{!}{%
    \begin{tabular}{@{}lccc@{}}
        \toprule
        &\multicolumn{3}{c}{Spoofing enhancement}\\\cmidrule(l){2-4}
        Anti-spoofing&None&wav2vec 2.0&HuBERT\\\midrule
        RawNet2&11.64&29.31&29.45\\
        wav2vec 2.0&8.28&20.05&19.51\\
        HuBERT&5.40&15.25&13.90\\
        WavLM&21.20&38.06&38.00\\
        WavLM+&10.76&29.20&28.10\\
        \bottomrule
    \end{tabular}%
    }}\hfill
    \subfloat[ASVspoof 2021 DF test set\label{tbl:main_results_part_2021DF}]{%
    \resizebox{0.32\linewidth}{!}{%
    \begin{tabular}{@{}lccc@{}}
        \toprule
        &\multicolumn{3}{c}{Spoofing enhancement}\\\cmidrule(l){2-4}
        Anti-spoofing&None&wav2vec 2.0&HuBERT\\\midrule
        RawNet2&24.47&49.76&48.24\\
        wav2vec 2.0&10.03&21.16&22.17\\
        HuBERT&15.07&38.40&38.16\\
        WavLM&14.78&42.30&42.48\\
        WavLM+&10.56&31.95&30.49\\
        \bottomrule
    \end{tabular}%
    }}
\end{table*}

\autoref{tbl:main_results_all} shows the EERs for the cases when the anti-spoofing models were trained using the whole ASVspoof 2019 LA training set.
Without spoofing enhancement models, the anti-spoofing models with the SSL frontend significantly outperformed the RawNet2 baseline on all the evaluation datasets, consistent with the previous study \cite{tak2022automatic}.
From the results on the 2019 LA test set (\autoref{tbl:main_results_all_2019LA}), the absolute EERs of RawNet2-based anti-spoofing were significantly increased by the spoofing enhancement, while those of the SSL-based anti-spoofing were hardly affected. This suggests that SSL models are not fooled by spoofing enhancement in clean conditions where the training and evaluation data do not diverge.
On the other hand, the results on the 2021 LA and DF test sets in Tables \ref{tbl:main_results_all_2021LA} and \ref{tbl:main_results_all_2021DF} show that not only RawNet2-based anti-spoofing but also SSL-based anti-spoofing were degraded by spoofing enhancement.
The performance gains that the defender side obtained by using SSL models were almost completely lost when the attacker side also used the SSL models.
Even so, since the RawNet2 baseline also suffers from performance degradation due to spoofing enhancement, the use of the SSL model for defense is recommended.
From the results, no particular performance degradation was observed when using the same pretrained model for the attacker and the defender, e.g., a \SI{16.57}{\percent} EER on the 2021 LA test set when both sides used wav2vec 2.0.
This means that it is a possible option to use publicly available SSL models on the Internet for defense.

\autoref{tbl:main_results_part} shows the EERs for the case when the anti-spoofing models were trained using the portion of the 2019 LA training set.
The trend in the results is the same as in \autoref{tbl:main_results_all}; spoofing enhancement did not affect the results for the clean condition (2019 LA) that much but degraded those for the realistic conditions (2021 LA and DF).

\begin{figure}[t]
    \centering
    \includegraphics[width=\linewidth]{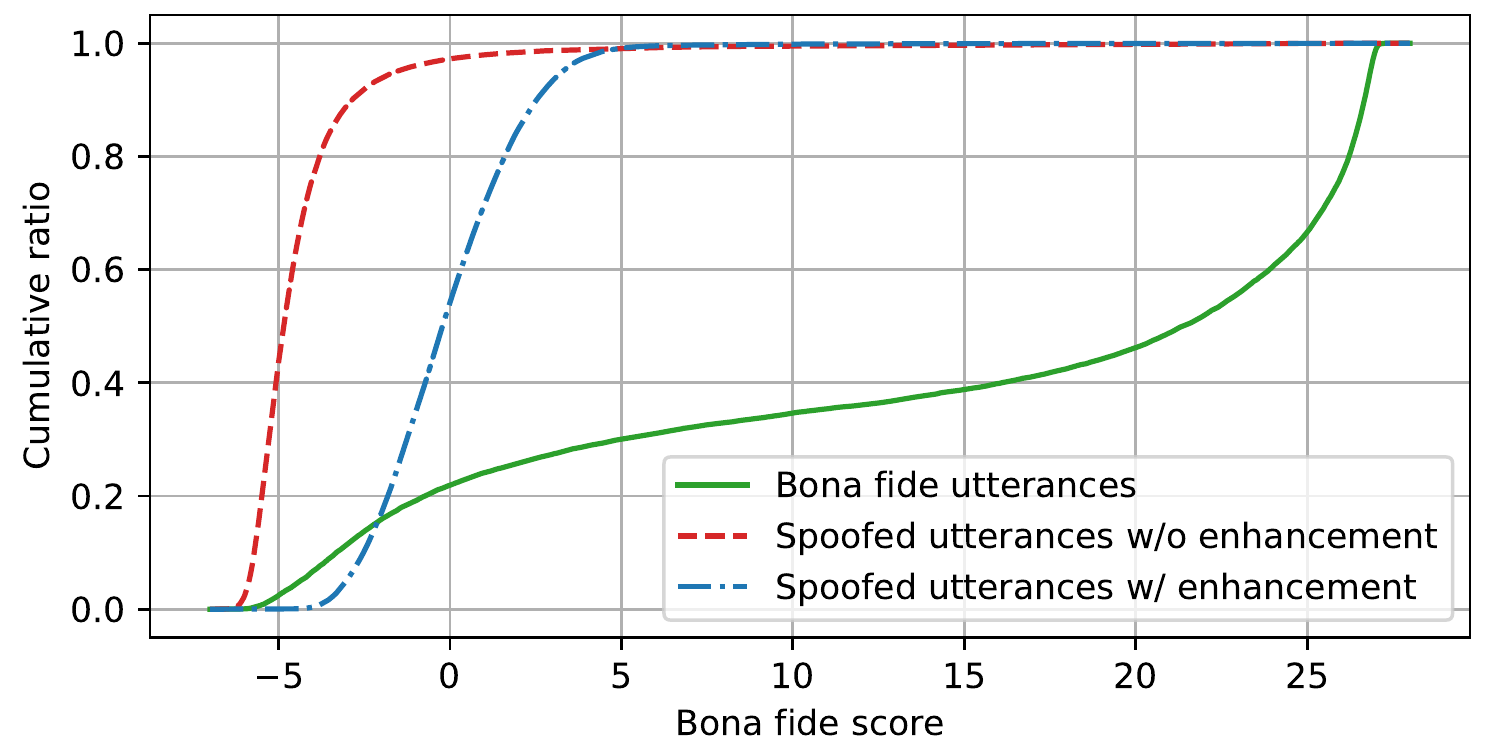}
    \caption{Cumulative distributions of bona fide score on ASVspoof 2021 LA test set. Anti-spoofing model and enhancement model are both based on wav2vec 2.0.}
    \label{fig:distribution}
\end{figure}

\autoref{fig:distribution} shows the cumulative distribution of the bona fide score, which was obtained as the logit of the bona fide class, for the 2021 LA set.
Both the anti-spoofing and spoofing enhancement models are based on wav2vec 2.0.
It is clearly observed that the spoofing enhancement shifted the distribution of the spoofed utterances to the right, \ie, increased the bona fide score.

\begin{figure}[t]
    \captionsetup[subfloat]{format=hang,singlelinecheck=false,justification=centering}
    \centering
    \includegraphics[width=0.32\linewidth]{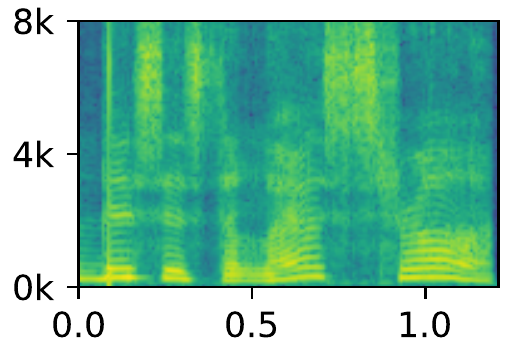}\hfill
    \includegraphics[width=0.32\linewidth]{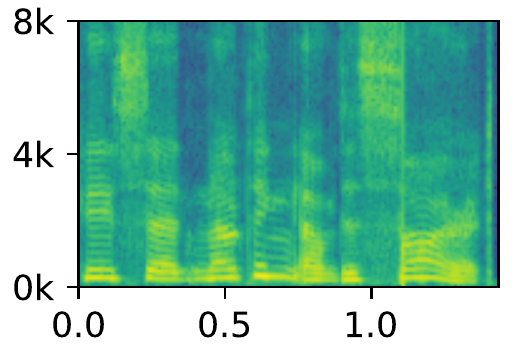}\hfill
    \includegraphics[width=0.32\linewidth]{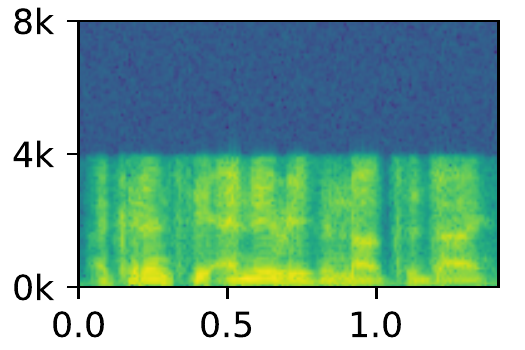}\\
    \subfloat[LA\_E\_8161161 \newline$(-5.7\rightarrow25.5)$]{\includegraphics[width=0.32\linewidth]{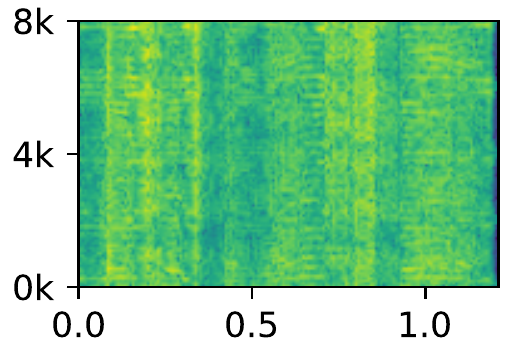}}\hfill
    \subfloat[LA\_E\_5770469 \newline $(-5.6\rightarrow23.7)$]{\includegraphics[width=0.32\linewidth]{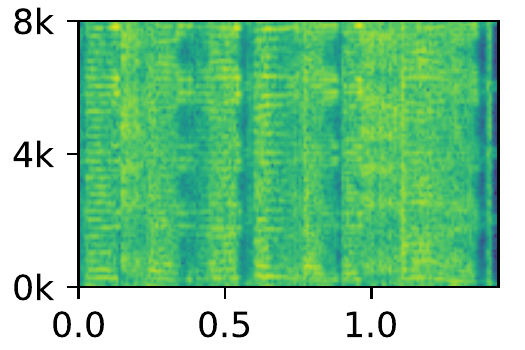}}\hfill
    \subfloat[LA\_E\_4822201 \newline $(-5.6\rightarrow21.1)$]{\includegraphics[width=0.32\linewidth]{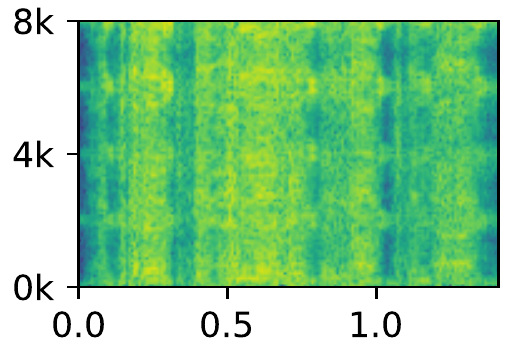}}
    \caption{Examples of spoofed utterances in ASVspoof 2021 LA test set before (top) and after (bottom) enhancement. Horizontal and vertical axes correspond to time and frequency, respectively. Anti-spoofing and enhancement models are both based on wav2vec 2.0. Values in brackets show how bona fide score was changed by enhancement.}
    \label{fig:examples}
\end{figure}

\autoref{fig:examples} shows examples of the original spoofed utterances and their conversions of which the bona fide scores were largely increased by the enhancement model based on wav2vec 2.0.
While spoofing enhancement makes utterances fool an anti-spoofing model, their harmonic structures are rarely preserved.
Spoofing enhancement that also preserves the naturalness of transformed utterances (like V2S \cite{nakamurav2s}) is left to future work.

\section{Conclusion}
In this study, we investigated whether the performance improvement of an anti-spoofing model obtained by using an SSL model is real, under the condition that the attacker also has access to the SSL model.
Experiments revealed that the attacker could also benefit from SSL models, thereby eliminating most of the benefits the defender gains from them.
We also found no significant EER degradation for the attacker side from using the same pretraining model as the defender side, indicating that both sides should simply use a stronger SSL model.
Future work will include a countermeasure for attackers in which pretrained SSL models are utilized.

\clearpage
\bibliographystyle{IEEEtran}
\bibliography{mybib}

\end{document}